\documentclass[10pt,conference,final,a4paper]{IEEEtran}
\usepackage{cite}      
\usepackage[caption=false, font=footnotesize]{subfig}

\ifCLASSINFOpdf
  \usepackage[pdftex]{graphicx}
  \DeclareGraphicsExtensions{.pdf}
\else
  \usepackage[dvips]{graphicx}
  \DeclareGraphicsExtensions{.eps}
\fi

\usepackage[latin1]{inputenc}
\usepackage{url}       
\usepackage{amsfonts}
\usepackage{amsmath}
\usepackage{times}
\usepackage{algpseudocode}
\usepackage{algorithm}
\usepackage{enumerate}
\usepackage{siunitx}
\usepackage{multirow}

\IEEEoverridecommandlockouts
\begin{document}
\title{\hspace*{-0.1mm}
On Optimization of Network-coded Scalable Multimedia Service Multicasting}
\author{\IEEEauthorblockN{Andrea Tassi\IEEEauthorrefmark{1}, Ioannis Chatzigeorgiou\IEEEauthorrefmark{1} and Dejan Vukobratovi\'c\IEEEauthorrefmark{3}}
\IEEEauthorblockA{\IEEEauthorrefmark{1}School of Computing and Communications, Lancaster University, United Kingdom}
\IEEEauthorblockA{\IEEEauthorrefmark{3}Department of Power, Electronics and Communication Engineering, University of Novi Sad, Serbia}\vspace{-4mm}\thanks{This work is part of the R2D2 project, which is supported by EPSRC under Grant EP/L006251/1. Collaboration of the authors was facilitated by COST Action IC1104 on Random Network Coding and Designs over $\mathrm{GF}(q)$.}}

\maketitle

\begin{abstract}
In the near future, the delivery of multimedia multicast services over next-generation networks is likely to become one of the main pillars of future cellular networks. In this extended abstract, we address the issue of efficiently multicasting layered video services by defining a novel optimization paradigm that is based on an Unequal Error Protection implementation of Random Linear Network Coding, and aims to ensure target service coverages by using a limited amount of radio resources.
\end{abstract}


\section{Introduction and Motivation}\label{sec:intro}
The technological evolution of communication devices has fuelled a surge in demand for new multimedia services over next-generation networks. Several standards for video coding and compression have been proposed. Among them, those that enable layered video streaming are gaining momentum. A layered video stream consists of one \emph{base layer} and multiple \emph{enhancement layers}. The base layer ensures a basic reconstruction quality, which progressively improves with the number of recovered enhancement layers. It is natural to adjust
the transmission of each video layer according to the user propagation conditions. In that way, users can eventually recover different sets of video layers, i.e., the same service at different quality levels.

Modern communication standards tackle the reliability issues of multicast communications by means of Application Level-Forward Error Correction (AL-FEC) schemes. However, these kind of codes are usually designed to be applied over large source messages. Hence, AL-FEC codes may lead to a \mbox{non-negligible} communication delay, which is an issue in the case of delay-sensitive multimedia services. That issue can be mitigated by using Unequal Error Protection implementations of Random Linear Network Coding (UEP-RLNC)~\cite{jsacTassi}.

Since the layers of a video stream have different importance levels, UEP-RLNC allows the transmitter to adjust the error protection capability of RLNC according to the importance level of the transmitted video layer. However, the capability of adjusting the protection level of each service layer is just one part of the general resource allocation issue, which is summarized in the following research questions: \textbf{[Q1]}~\emph{How to define an optimization framework that can jointly optimize both transmission parameters and \mbox{UEP-RLNC} parameters?} \textbf{[Q2]}~\emph{How can a service provider allocate radio resources such that an existing Service Level Agreement towards users is not violated?} In the following sections, we will we will propose an answer to those questions.

\section{System Model and Proposed Optimization}\label{sec:sm}
Consider an Orthogonal Frequency-Division Multiple Access (OFDMA) cellular system composed by a Base Station (BS) and $U$ users. A layered data stream is transmitted to the users over $C$ orthogonal broadcast erasure subchannels. Since the service is encoded as in the RLNC principle, each subchannel conveys a stream of \textit{coded packets}. Coded packets transmitted over the same subchannel adopt the same Modulation and Coding Scheme\footnote{If $m^\prime$ and $m^{\prime\prime}$ are indexes of two MCSs and $m^\prime \leq m^{\prime\prime}$, the MCS described by $m^{\prime\prime}$ either uses a higher modulation order or reduced error-correcting capability than the MCS represented by $m^{\prime}$.} (MCS).  Subchannel $c$ is modeled as a \mbox{frequency $\times$ time} structure, which spans a certain bandwidth and multiple OFDM symbols. Each coded packet transmitted over $c$ spans a fixed number of OFDM symbols and the same bandwidth of $c$. The bit length of the coded packets is fixed, while the subchannel bandwidth is adapted -- in accordance with the MCS $m_c$ adopted by $c$ -- to fit the bit length of the coded packets.

\begin{figure}[tbd]
\centering
\includegraphics[width=0.85\columnwidth]{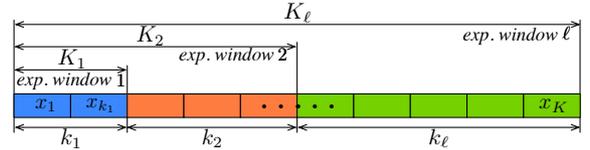}
\vspace{-2mm}\caption{Considered layered source message.}
\label{fig.msg}
\end{figure}

We model a layered service as a sequence of source messages. Each message $\mathbf{x}$ consists of $K$ source packets. The elements of $\mathbf{x}$ are grouped into $L$ \emph{layers}, as in Fig.~\ref{fig.msg}. We define the user Quality-of-Service (QoS) level as the number of \emph{consecutive} layers that can be recovered, starting from the first layer. According to the UEP-RLNC presented in~\cite[Sec.~II.B]{jsacTassi}, for each expanding window $\mathbf{x}_{1:\ell}$, a stream of coded packets $\{y_1, \ldots, y_{N_\ell}\}$ is generated, where $y_j = \sum_{i = 1}^{K_\ell} g_{j,i}\cdot x_i$ and $g_{j,i}$ is randomly selected over $\mathrm{GF}(q)$ of size $q$. We propose the UEP Resource Allocation Model (UEP-RAM), which jointly optimises the number $N_\ell$ of coded packet transmissions associated with each expanding window $\ell$, and the MCS $m_\ell$ used to transmit them. For simplicity, we assume that coded packets associated with different expanding windows are kept separated and transmitted over different subchannels.

We define the indicator variable $\delta_{u,\ell}$ associated with user $u$ and QoS level $\ell$ such that $\delta_{u,\ell}\!=\!1$ if the first $\ell$ service layers are recovered with a probability of at least $\Hat{Q}$; otherwise, $\delta_{u,\ell} = 0$. User $u$ will recover the first $\ell$ service layers if it successfully recovers the $\ell$-th expanding window or \textit{any} of the expanding windows with index greater than $\ell$. Thus, we define $\delta_{u,\ell} = I\left(\bigvee_{i = \ell}^{L} \textrm{P}_{u,i}(N_1, \ldots, N_i, m_1, \ldots, m_i) \!\geq\! \Hat{Q}\right)$, where $I(\cdot)$ is the indicator function\footnote{If $s$ is true, $I(s) = 1$, otherwise $I(s) = 0$.}, and $\textrm{P}_{u,i}(\cdot)$ is the probability that $u$ collects $K_i$ linearly independent coded packets associated with expanding windows with indexes $1, \ldots, i$~\cite[Eq.~(10)]{jsacTassi}.

We define the \emph{system profit} $\sum_{u = 1}^U\sum_{\ell= 1}^L \delta_{u,\ell}$ as the number of video layers that any of the $U$ users can recover, while the \emph{system cost} is the total number of coded packet transmissions $\sum_{\ell = 1}^L N_\ell$. Inspired by a fundamental economics principle, in order to optimize the profit while keeping the cost low, the proposed UEP-RAM maximizes the \emph{profit-cost ratio} as~\cite{iccTassi}
\begin{align}
	\text{(UEP-RAM)} &  \quad  \mathop{\mathop{\textrm{maximize}}_{m_1, \ldots, m_L}}_{N_1, \ldots, N_L} \,\,  \sum_{u = 1}^U\sum_{\ell= 1}^L \delta_{u,\ell} \Bigg/ \sum_{\ell = 1}^L N_\ell \label{UEP.of}\\
   \textrm{subject to} &   \quad \sum_{u = 1}^U \delta_{u,\ell} \geq U \, \Hat{t}_\ell \quad\quad\quad\,\text{$\ell= 1, \ldots, L$} &\label{UEP.c1}\\
                     &   \quad 0 \leq N_\ell \leq \Hat{N}_\ell \quad\quad\quad\quad\text{$\ell= 1, \ldots, L$.}&\label{UEP.c2}
\end{align}
Constraint~\eqref{UEP.c1} ensures that the fraction of users recovering the first $\ell$ video layers shall not be smaller than $\Hat{t}_\ell$. Constraint~\eqref{UEP.c2} imposes that the transmission of expanding window $\ell$ shall not require more than $\Hat{N}_\ell$ coded packet transmissions. Since MCSs and coded packet transmissions per expanding window are optimized altogether, UEP-RAM is the answer to [Q1]. Furthermore, the constraint set~\eqref{UEP.c1}-\eqref{UEP.c2} ensures that predetermined fractions of users recover given subsets of service layers by a certain deadline. That represents the requirements of a SLA to be honoured, hence, [Q2] has also been addressed.

\begin{figure}[t]
\centering
\includegraphics[width=0.99\columnwidth]{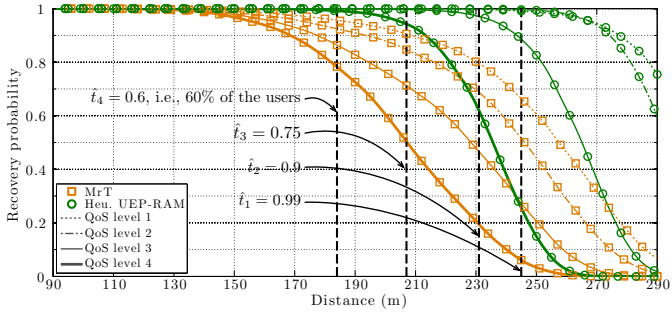}
\vspace{-7.5mm}\caption{Probability of achieving the QoS level $\ell$ vs. the distance from the centre of the cell, for MrT and heuristic UEP-RAM (``Heu. UEP-RAM''), \mbox{$\ell = 1, \ldots, L$}, in the Single Cell scenario. Vertical dashed lines denote the target coverage associated with each QoS level.}\vspace{-1.5mm}
\label{fig.sc}
\end{figure}

In spite of the apparent complexity of UEP-RAM, it is possible to derive an efficient heuristic strategy by following the same train of thoughts as in~\cite{jsacTassi}. In particular, the proposed heuristic procedure comprises two steps: (i) the optimization of $m_1, \ldots, m_L$, and (ii) the optimization of $N_1, \ldots, N_\ell$. That heuristic can produce good quality solutions in a finite number of steps~\cite{iccTassi}.

\vspace{-2mm}\section{Numerical Results and Discussion}\label{sec:nr}
We consider a Long Term Evolution-Advanced (LTE-A) network composed of $19$ base stations, each of which controls three hexagonal cell sectors. We compared the proposed heuristic UEP-RAM against a common Multi-rate Transmission Strategy (MrT)~\cite[Eq.~(31)-(32)]{jsacTassi}, which (i) does not employ RLNC or AL-FEC error protection strategies, and \mbox{(ii) aims} at maximizing the average user QoS by optimising the MCSs used to transmit each video layer.

First, we consider the \emph{Single Cell (SC) scenario}, where $18$ interfering base stations are organized in two concentric rings centred on the base station multicasting a 4-layer H.264/SVC video stream~\cite{6025326}. Users ($U = 80$) are regularly placed on the radial line representing the symmetry axis of a cell sector of the central cell. Fig.~\ref{fig.sc} shows the probability of recovering the first $\ell$ video layers as a function of the distance from the base station in the centre of the cell. We observe that the proposed UEP-RAM comfortably meets the service coverage constraints. In addition, UEP-RAM ensures a service coverage which is considerably greater than that of MrT.

\begin{figure}[t!]
\centering
\includegraphics[width=0.63\columnwidth]{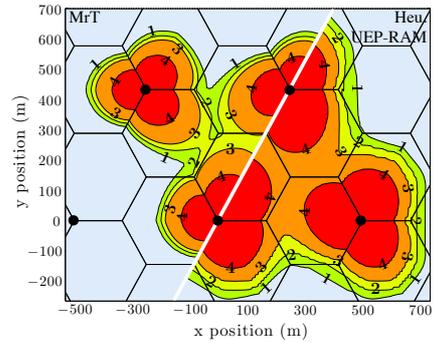}
\vspace{-3mm}\caption{QoS levels provided by the MrT and heuristic UEP-RAM (``Heu. UEP-RAM'') strategies (black circles represent the base stations), in the SFN scenario.}\vspace{-1mm}
\label{fig.sfn}
\end{figure}

We also considered the \emph{Single Frequency Network (SFN) scenario}, where $4$ base stations, surrounded by the remaining $15$ interfering base stations, realize a LTE-A SFN~\cite{sesia2011lte}. The SFN multicasts the aforementioned video stream to users \mbox{($U = 1700$)} placed at the vertices of a regular square grid placed on the playground. Also in this case, coverage constraints are met. Fig.~\ref{fig.sfn} also provides a visual comparison of the coverage areas offered by MrT and UEP-RAM. Coloured regions depict areas where users can achieve a specific QoS level. Once more, we observe that UEP-RAM ensures a service coverage which is considerably greater than that of MrT.

\bibliographystyle{IEEEtran}
\bibliography{IEEEabrv,biblio}

\end{document}